\documentclass[final,3p,times]{elsarticle}
\usepackage{amssymb}
\usepackage{natbib}
\biboptions{sort&compress}
\usepackage{epsfig,epsf,graphicx,amsmath,graphics,color}
\journal{Nuclear Physics A}
\begin{document}
\begin{frontmatter}
\title{Theoretical analysis of $^8$Li + $^{208}$Pb reaction 
and the critical angular momentum for complete fusion}
\author[unisa,ift]{B. Mukeru}
\author[unisa]{M. L. Lekala}
\author[uff]{J. Lubian}
\author[ift]{Lauro Tomio}
\ead{lauro.tomio@unesp.br}
\address[unisa]
{Department of Physics, University of South Africa, PO Box 392, Pretoria 0003, South Africa.}
\address[ift]
{Instituto de F\'isica T\'eorica, Universidade Estadual Paulista, 01140-070 S\~ao Paulo, SP, Brazil.}
\address[uff]
{Instituto de F\'isica, Universidade Federal Fluminense, Avenida Litoranea s/n, Gragoat\'a, Niter\'oi, RJ, 24210-340, Brazil.} 

\begin{abstract}
 In a theoretical approach, the complete and incomplete fusions are investigated by considering the 
$^8{\rm Li}+{}^{208}{\rm Pb}$ reaction.  By decreasing the projectile ground-state binding energy  $\varepsilon_b$ 
from its known experimental value, the complete fusion is shown to have insignificant dependence on such variations,  
whereas the incomplete fusion strongly depends on that. 
The complete and incomplete fusion cross sections are calculated by using a combination of both continuum-discretized 
coupled-channel and sum-rule models. To this end, an incident-energy dependent cut-off angular momentum $L_c$ is 
first obtained by using the available complete fusion experimental data,  
within an approach which is extended to model results obtained for other incident-energies.
An approximated fitted expression linking $L_c$ to the well-known critical value $L_{\rm crit}$ derived by  
Wilczy\'nski [Nucl. Phys. A 216 (1973) 386] suggests a generalization of the corresponding sum-rule model
to energies around and below the Coulomb barrier.\\
\end{abstract}

\end{frontmatter}

\date{\today}

\section{Introduction}
\label{intro}

The studies related to fusion reactions induced by loosely-bound projectiles, and their corresponding break up possibilities, 
are currently among the hottest subjects in Nuclear Physics 
\cite{Canto10,Torres20,2006PLB-Gomes,Yu10,2009NPACanto,Gomes21,2012NPAPaes,Gomes10,Wang10,2014Carlson,2014Capel,Hu10,2015Boselli,2015Lay,Kundu10,Guo10,Parkar20,2016FBSGomes,Raj20,Kumar50,Singh10,Torres10,Parkar30,Mukeru10,Kolinger10,Cook20,Cook21}. 
Two kinds of fusions have emerged from these investigations, namely the complete fusion (CF), and the incomplete fusion (ICF),
whose sum amounts to the total fusion (${\rm TF}={\rm CF}+{\rm ICF}$). Given the low projectile binding energy,
there is a high probability that the latter breaks up into two or more fragments before reaching the absorption region. 
In this case, all the fragments may be absorbed by the target, leading to the 
complete fusion. On the other hand, the target may absorb some but not all the fragments, while the scattered ones
fly on the outgoing trajectory, 
leading to incomplete fusion. Many other phenomena such that transfer, delayed breakup may occur during such collisions,
which are not discussed in this work. Nevertheless, one should point out that if the projectile breaks up when is moving 
apart of the target, this so-called delayed breakup will affect the reaction cross section, but not the fusion cross 
section~\cite{Canto10}. 
Complete fusion is regarded as the absorption of the whole charge of the projectile, and not necessarily its whole mass, 
while incomplete fusion refers to the absorption of only a part of the projectile charge~\cite{Canto10,Gomes10,Parkar20,Cook20}. 
On the other hand, complete fusion is also interpreted as the absorption from bound states, with   
the  incomplete fusion as the absorption from breakup states~\cite{Torres20}. 
Yet, another model being used to study complete and incomplete fusions is the sum-rule model presented in Refs.~\cite{Wik20,Wik10,1982Wik},   in which the idea of partial statistical equilibrium is combined with the 
generalized concept of critical angular momentum. 
According to this model, the CF process occurs at lower angular momenta ($L\le L_{\rm crit}$),
where $L_{\rm crit}$ is a critical angular momentum, that separates both complete and incomplete processes. In other words, the
incomplete fusion process occurs at higher angular momenta ($L>L_{\rm crit}$). Complete and incomplete fusion cross sections 
obtained by using this model are reported in several works (see, for example, Refs.\cite{Ina20,Ger20,Trau20,K100,LR10,
  Yadav10,Vijay10,Muntazir10,Sabir10}). 
 However, this approach has being applied only for energies well above the Coulomb barrier, with $L_{\rm crit}$ being 
independent of the incident energy. 
In view of that, an interesting study could be to verify whether such an approach can be 
extended to lower incident energies, around the Coulomb barrier,  where in fact the incomplete fusion is important and 
both fusion processes are expected to have strong dependence on the incident energy. 

Although the complete fusion process have been extensively studied, the full understanding of its suppression due to 
the projectile breakup  is yet to be fully understood.  For example, while it is widely believed that the suppression of 
complete fusion has a stronger dependence on the projectile breakup threshold (see, for instance, Refs.~\cite{Wang10,Rath100} 
and references therein), recent experimental results in $^{7,8}{\rm Li}+{}^{209}{\rm Bi}$ reactions have suggested 
charge clustering rather than weak projectile binding (i.e., breakup prior to reaching the fusion barrier) as the crucial factor in 
complete fusion suppression~\cite{Cook20,Cook21}. It is further asserted in these references that
 weak projectile binding energy leads to incomplete fusion enhancement, while strong charge clustering 
leads to complete fusion suppression. These fascinating results call for further investigation in the role of the projectile 
breakup on the complete fusion process.  Among other approaches, this can be achieved by artificially varying 
the projectile ground-state binding energy. One of the advantages of such an approach is that, it keeps the projectile mass 
and charge unchanged, thus minimizing their effects. In fact, this procedure was adopted in Ref.~\cite{Lei100} for 
$^{6,7}{\rm Li}$ projectiles, providing numerical support to results reported in Ref.~\cite{Cook20}. 
The artificial variation of the ground-state binding energy in that reference, revealed another interesting aspect:
in the case of $^7{\rm Li}$ projectile, for binding energies weaker than the experimental values, the complete fusion cross 
section displays an insignificant dependence on this energy variation, while the opposite was observed when considering the
$^6{\rm Li}$ projectile (see Fig. 3 of Ref.~\cite{Lei100}).
For energies weaker than the experimental breakup threshold, one is representing a clear case of projectile 
breakup prior to reaching the fusion barrier. Therefore, to some extend, these results clarify the fact that weak binding 
energy alone is not a sufficient condition to warrant the suppression of the complete fusion cross section, in accordance
to Refs.\cite{Cook20,Cook21}. Since the complete fusion process can also be regarded as the absorption of the whole 
projectile charge, one would as well expect this process to exhibit an insignificant dependence on variations of the 
ground-state biding energy in the case of $^8$Li ($^7$Li+$n$), regardless the target mass. 
    
By considering the $^7{\rm Li}$ results~\cite{Lei100} together with the sum-rule model of Ref.~\cite{Wik20}, it
can be argued that
the total fusion cross section would exhibit an insignificant dependence on the variation of the ground-state binding 
energy in the case the angular momentum $L$ is smaller than a cutoff value $L_c$, such that for the 
angular momentum window $L\le L_{c}$, the complete fusion is the most dominant process. 
For higher angular momenta where the incomplete fusion is more important than its complete counterpart, the total 
fusion would be expected to strongly depend on the variation of the binding energy. This assessment is further justified 
by the fact that complete fusion is derived from bound-state absorptions, whereas incomplete fusion derives from
 breakup state absorptions. 
 Such study would serve not only as a further numerical proof of the conclusions
drawn in Refs.\cite{Cook20,Cook21}, but to some extend can be useful to establish a connection
between different definitions for complete and incomplete fusion processes.

In this paper, we study total, complete and incomplete fusion processes  in the breakup of $^8{\rm Li}$
  projectile on a lead target. We are particularly interested in investigating the dependence of the complete fusion cross section
  on the variation of the ground-state binding energy below the experimental value.  As a first step, we calculate the 
  total fusion cross section  by means of the of the Continuum-Discretized Coupled-Channel (CDCC) method \cite{Aust100}.
  Next, the complete fusion cross section is obtained from the calculated total fusion cross section, 
  by following the sum-rule model, as given in  Ref.~\cite{Wik20}. 
  Without diving into the details about the model, the interested reader can also found some fundamentals and 
  relevant discussion in section 4 of Ref.~\cite{1982Wik}.
  In this case, we sum up the  angular momentum distribution total fusion cross sections from zero up to some upper limit
  $L_{c}$, identifying the remaining part of the sum as corresponding to incomplete fusion cross section. Therefore,
  the crucial ingredient in this process is the angular momentum cutoff $L_c$, which is first determined by   
  using the available complete fusion experimental data, as it will be shown. Next, this procedure is extended to regions 
  where experimental data are not available, by considering complete fusion model results.
  This parameter $L_c$ emerges naturally as dependent on the incident energy, as well as on the projectile binding energy. 
For the reaction under study,
 an  attempt is made in establishing a relationship between the energy-dependent $L_c$ and the $L_{\rm crit}$ of  
 Ref.~\cite{Wik20}.  
 Once $L_c$ is obtained,  our next step will be to analyze  the dependence of the total fusion cross 
 section on the variation of the ground-state binding energy below the experimental value, for $L\le L_c$ and for $L>L_c$; 
 leading to the study the dependence of the complete and incomplete fusion cross sections on such  binding energy variation.
Apart from the $^8{\rm Li}$ experimental ground-state energy $\varepsilon_b=2.033\,{\rm MeV}$~\cite{Nut10}, 
three more values are arbitrarily considered below the experimental one: $\varepsilon_b=$1.5, 1.0 and 0.5 MeV, obtained 
by adjusting the depths of the central and spin-orbit coupling components of the  Woods-Saxon potential used for the 
bound as well as the continuum wave functions.

In the next, we organize the presentation of this paper as follows:  
the CDCC formalism is briefly described in Section~\ref{descr}, with details 
of the calculations presented in Section \ref{calc}.  The results are presented and discussed in Section~\ref{results}, with our 
conclusions given in Section~\ref{conclusion}.

\section{Projectile-target basic CDCC formalism}
\label{descr}
Following Ref.\cite{Aust100} for a system having a target with a weakly-bound core-plus-neutron projectile, the 
corresponding three-body Schr\"odinger equation is transformed into the CDCC differential equation after an expansion of the 
three-body wave function on a complete basis of bound and continuum states of the projectile~\cite{fresco}.
 With ${\bf R}$ being the vector
position of the target in relation to the projectile center-of-mass, the corresponding radial coupled differential equations for the
wave-function components $\chi_{\alpha}^{LJ}(R)$ are given by
{\begin{eqnarray}\label{coupled}
\left[-\frac{{ \hbar}^2}{2\mu_{pt}}\bigg(\frac{d^2}{dR^2}-\frac{L(L+1)}{R^2}\bigg)
+U_{\alpha\alpha}^{LJ}(R)\right]\chi_{\alpha}^{LJ}(R)
%\nonumber\\
+\sum_{\alpha\ne\alpha'}U_{\alpha\alpha'}^{LL'J}(R)\chi_{\alpha'}^{L'J}(R)=(E-\varepsilon_\alpha)\chi_{\alpha}^{LJ}(R),
\end{eqnarray}}
where $\mu_{pt}$ is the projectile-target reduced mass, with the quantum numbers $L$ and $J$ being 
identified, respectively, with the orbital and total angular momentum (which, in the following, will be given in units of $\hbar$).
The other relevant quantum numbers to describe the states of the projectile are represented by $\alpha\equiv \{i, \ell, s, j\}$ 
($i=0, 1, 2,\ldots, N_b$, with $N_b=$ number of bins), with  $\varepsilon_{\alpha}$ being the projectile bin energies.
By having $j\equiv l+s$, $l$ is the relative angular momentum between $^7$Li and the neutron, 
with $s$ being the spin of the neutron, considering that the 
interaction does not depend on the spin of the core.
$U_{\alpha\alpha'}^{LL'J}$ are the 
potential matrix elements, which are defined by 
\begin{eqnarray}\label{potmatr}
U_{\alpha\alpha'}^{LL'J}(R)=\langle\mathcal{F}_{\alpha L}({\bf r},\hat R)|V_{pt}({\bf R},{\bf r})|\mathcal{F}_{\alpha' L'}({\bf r},\hat R)\rangle,
\end{eqnarray}
where
\begin{eqnarray}\label{channel}
  \mathcal{F}_{\alpha L}({\bf r},\hat R)=[i^L\Phi_{\alpha}({\bf r})\otimes Y_L^{\Lambda}(\hat R)]_{JM},
\end{eqnarray}
with the function $\Phi_{\alpha}({\bf r})$  containing the bound and discretized bin wave functions of the projectile.  The
potential $V_{pt}({\bf R},{\bf r})$ in Eq.(\ref{potmatr}) is a sum of core($^7{\rm Li}$)-target and neutron-target potentials, 
such that
\begin{eqnarray}\label{pot}
V_{pt}({\bf R},{\bf r})=V_{ct}({\bf R}_{ct})+V_{nt}({\bf R}_{nt}),\nonumber\\
  [2mm]
  {\bf R}_{ct}={\bf R}+\frac{1}{8}{\bf r},\quad  {\bf R}_{nt}={\bf R}-\frac{7}{8}{\bf r}.
\end{eqnarray}
It contains both Coulomb and nuclear components to account for the total breakup. The imaginary part of the projectile-target nuclear
potential, $W_{pt}({\bf R},{\bf r})=W_{ct}({\bf R}_{ct})+W_{nt}({\bf R}_{nt})$ is responsible for the absorption of the projectile by the target.
Therefore, the coupling matrix elements corresponding to the aborption is given by
\begin{eqnarray}\label{CMI}
  W_{\alpha\alpha'}(R)=\langle\mathcal{F}_{\alpha L}({\bf r},\hat R)|W_{ct}({\bf R}_{ct})+W_{nt}({\bf R}_{nt})|\mathcal{F}_{\alpha' L'}({\bf r},\hat R)\rangle.
\end{eqnarray}
The total fusion cross section (absorption cross section) is then obtained as the expectation given by~\cite{Kolinger10,Ha20}
  {\small
\begin{eqnarray}\label{TA}
  \sigma_{\rm TF}&=&\sum_{L=0}^{L_{\rm max}} \sigma_{\rm TF}^{(L)}\equiv\sum_{L=0}^{L_{\rm max}}
\left[  \frac{2\mu_{pt}}{\hbar^2 K_0}(2L+1)\sum_{\alpha\alpha'}
  \langle \chi_{\alpha}^{LJ}(R)|W_{\alpha\alpha'}(R)|\chi_{\alpha'}^{L'J}(R)\rangle\right],  
  \end{eqnarray}}
where
$\chi_{\alpha'}^{LJ}(R)$ is the full-radial wave function, and 
and $K_0$ is the projectile-target relative wave number in the incident channel.
The complete fusion cross section can be directly obtained from Eq.(\ref{TA}), according to the sum-rule model as follows
\begin{eqnarray}\label{SUMR}
\sigma_{\rm CF}=\sum_{L=0}^{L_c}\sigma_{\rm TF}^{(L)},
\end{eqnarray}
where $L_c$ is the angular momentum cutoff. The incomplete fusion is then given by 
\begin{eqnarray}\label{ICF}
\sigma_{\rm ICF}=\sigma_{\rm TF}-\sigma_{\rm CF}=\sum_{L>L_c}^{L_{\rm max}}\sigma_{\rm TF}^{(L)}.
\end{eqnarray}

\section{Calculation details}
 \label{calc}
\begin{figure}[!!h]
\begin{center}
  \resizebox{8cm}{!}{\includegraphics{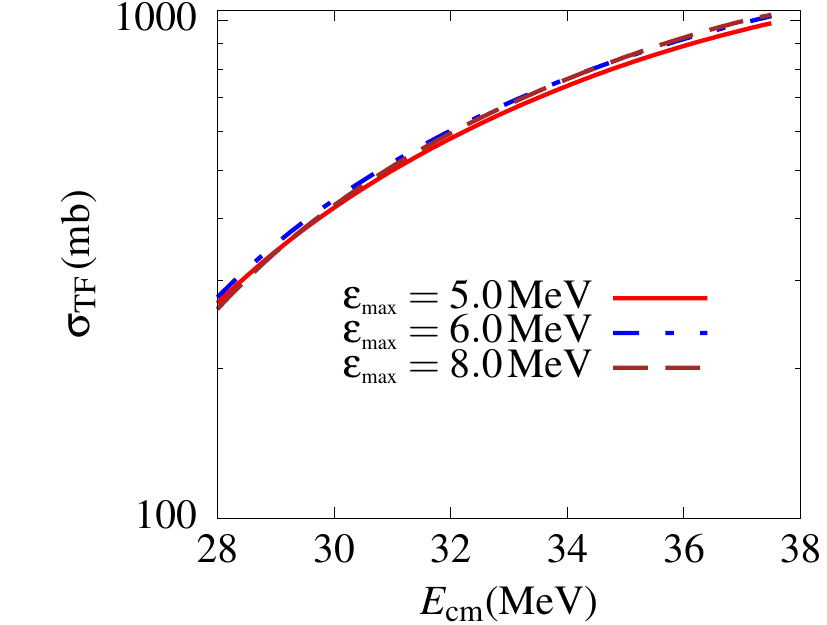}}
\end{center}
\vspace{-0.5cm}
\caption{\label{fig:1}  {
Convergence of $^8{\rm Li}+{}^{208}{\rm Pb}$ total fusion cross section results (with $n-^7$Li binding 2.03 MeV), 
as the maximum bin energy $\varepsilon_{\rm max}$ varies from 5 to 8 MeV.}}
\end{figure}

 In the CDCC formalism, the projectile energies and wave functions are crucial inputs in the coupling matrix elements.
 In the present work, the 
 projectile being considered is the $^8{\rm Li}$, which is known to be a $n$-${}^7{\rm Li}$ bound system, having
 a ground-state binding energy $\varepsilon_b=2.033\,{\rm MeV}$, with $j^{\pi}=2^+$. In a single-particle
 (or shell-model) picture, this ground state can be interpreted 
 as a valence neutron in a ($\ell=1$, $n=1$) single-particle configuration. 
This nucleus exhibits also a $j^{\pi}=1^+$ first excited state with energy $\varepsilon_{\rm ex}=0.98\,{\rm MeV}$ \cite{Nut10}. 
The ground and excited states, as well as the continuum wave functions, are  first obtained 
by using the parametrization presented in Ref.~\cite{Moro200} for the usual Woods-Saxon plus spin-orbit (SO) nuclear potential,  
as given in Refs.~\cite{1995Martel,1996Esbensen}, with $V_{\rm SO}=4.89\,{\rm MeV\cdot fm^2}$, $r_{\rm SO}=r_0=1.25\,{\rm fm}$ 
and $a_{\rm SO}=a_0=0.52\,{\rm fm}$.
The depth of the central potential, $V_0$, which is adjusted to take into account the ground, excited states and continuum wave
functions, is also used as a parameter for tuning the ground-state binding energies being considered in the present study.
The global parametrization of Akyuz-Winther\cite{Winter20} is used to obtain the parameters of real parts of the 
$^7{\rm Li}$-target and $n$-target optical potentials. 
  As for the imaginary parts, we follow Ref.~\cite{Torres20} and consider
short-ranged imaginary potentials with parameters $W=50\,{\rm MeV}$, $r_w=1.0\,{\rm fm}$ and $a_w=0.10\,{\rm fm}$. 
These parameters are used for both $^7{\rm Li}$-target and $n$-target imaginary potentials. 
The short-ranged nature of these potentials implies that they quickly vanish beyond $r_w(A_c^{1/3}+A_t^{1/3})$ for 
the core-target system, and $r_wA_t^{1/3}$ for the $n$-target system, where $A_c=7$ and $A_t=208$ are the core and 
target atomic mass numbers, respectively. We also verify that, as long as this potential is well inside the Coulomb barrier, the 
results have a small dependence on such parameters.  
For the numerical calculations, we consider a maximum bin energy $\varepsilon_{\rm max}=$8\,MeV. The
 $[0,\varepsilon_{\rm max}]$ interval is discretized into bins of width $\Delta\varepsilon=0.5\,{\rm MeV}$ for $s-$ and $p-$waves, 
 $\Delta\varepsilon=1.5\,{\rm MeV}$ for $d-$waves, and $\Delta\varepsilon=2.0\,{\rm MeV}$ for higher partial waves.
The angular momentum between $^7{\rm Li}$ and the neutron is truncated at $\ell_{\rm max}=4$, with the maximum 
matching radius for the bin integration over the coordinate being  
$r_{\rm max}=70\,{\rm fm}$. The $^7{\rm Li}$-target and $n$-target potentials are expanded into potential multipoles 
of maximum $\lambda_{\rm max}=4$.
For the integration of the coupled differential equations over relative centre-of-mass distances, the matching radius is 
$R_{\rm max}=600\,{\rm fm}$, with the angular momentum of the relative centre-of-mass motion truncated at 
$L_{\rm max}=500$. 
These parameters are selected in accordance with the convergence requirements, for the considered $^8{\rm Li}$
binding energies. This convergence is exemplified in Fig.~\ref{fig:1}, for results obtained for the total fusion cross section
as a function of the incident center-of-mass (c.m.) energy $E_{\rm c.m}$, assuming different maximum bin energies for 
$\varepsilon_b=2.03\,{\rm MeV}$, which are indicating that $\varepsilon_{\rm max}=$8\,MeV is quite enough to insure 
converged results in all the cases.
\begin{figure}[!!h]
\begin{center}
  \resizebox{8cm}{!}{\includegraphics{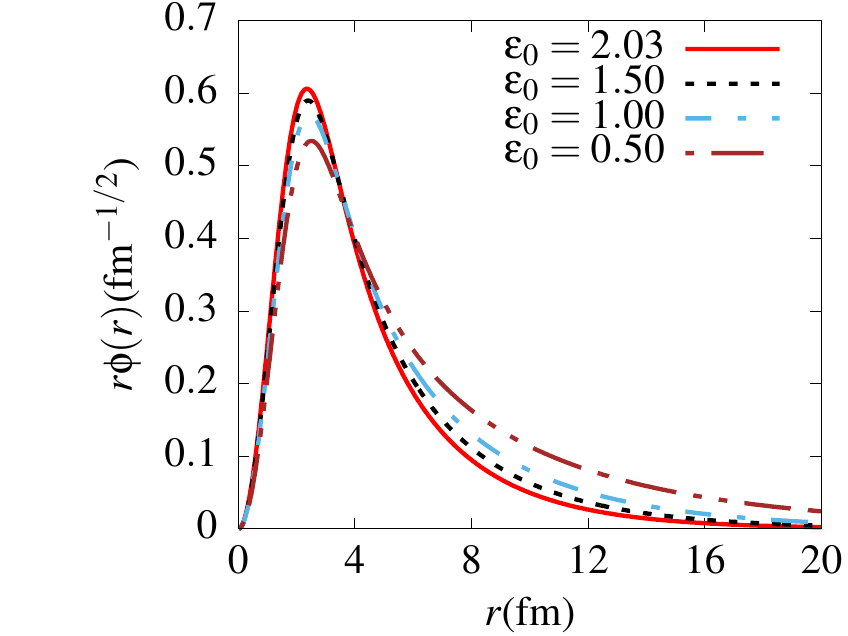}}
\end{center}\vspace{-0.5cm}
\caption{\label{fig:2}  Ground-state radial wave functions, given by $r \phi({ r})$ (in fm$^{-1/2}$), for different $n-^7$Li binding energies 
$\varepsilon_b$ (in MeV), as indicated inside the frame.} 
\end{figure}

\section{Results and Discussion}
\label{results}

\begin{figure}[t]
	\begin{center}
		\resizebox{8cm}{!}{\includegraphics{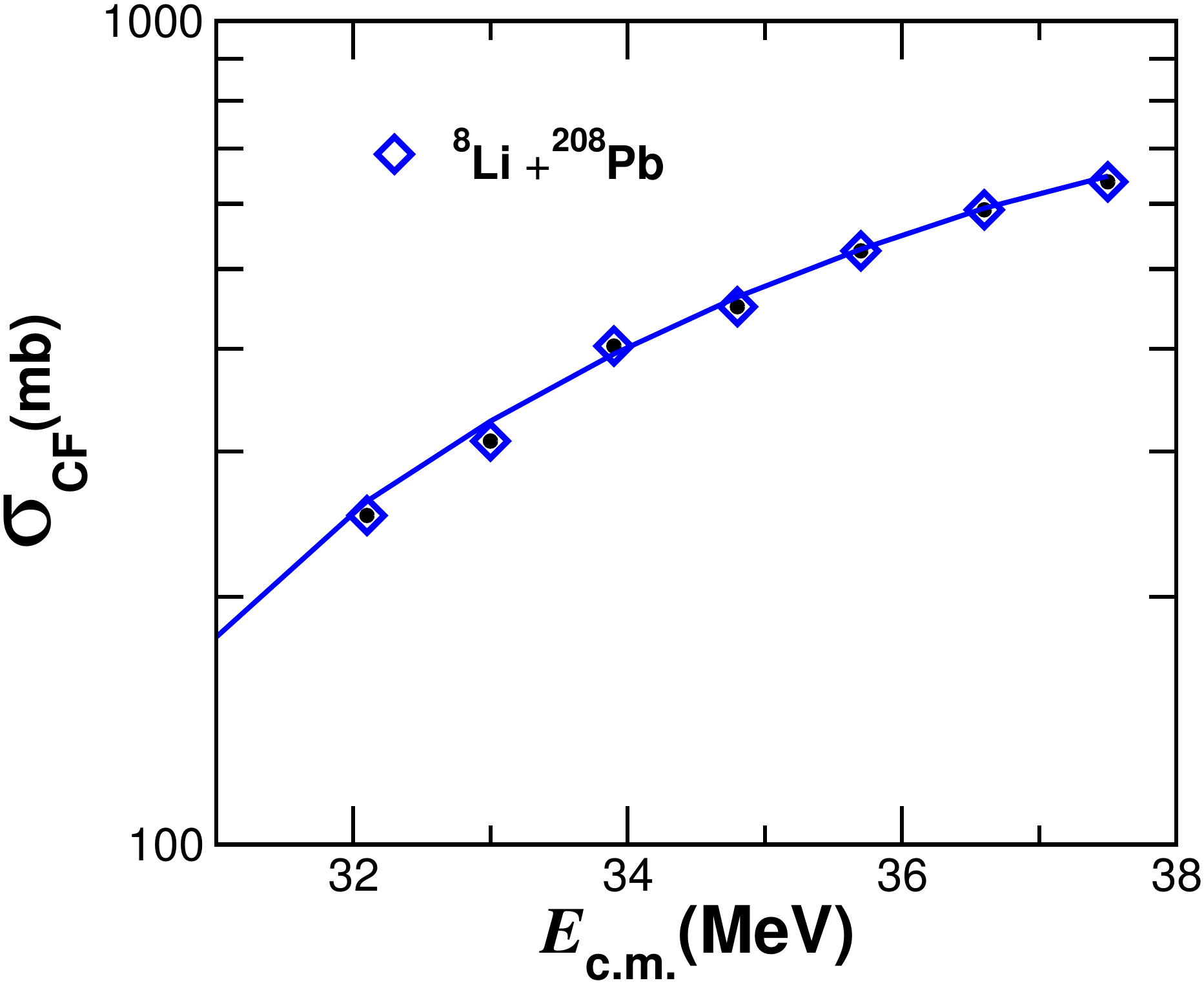}}
	\end{center}\vspace{-.5cm}
	\caption{\label{fig:3} Complete fusion cross sections (solid line) calculated by using Eq.(\ref{SUMR}),
	with $L_c$ as given in Table 1.
          The experimental data (symbols) are from Table 1 of Ref.\cite{Aguilera10}, with the corresponding error
          bars (absolute values $\le 13$ mb) being approximately within the size of the symbols.
	}
\end{figure}

The ground-state radial wave functions for $n$-${}^7{\rm Li}$ system, given by $r \phi({ r})$, are displayed in
Fig.~\ref{fig:2} by considering different 
binding energies, which are obtained by varying the strength of the corresponding nuclear potential. 
As one can notice from this figure, as the binding energy decreases, it appears that the magnitude of the
wave functions also decrease in the inner part 
($r\le 4\,{\rm fm}$), while the corresponding densities are extended to larger distances, due to the longer tails of the 
ground-state wave functions. Therefore, it is of interest to verify how this behavior will affect the
  complete and incomplete fusion cross sections.
  As anticipated in the introduction, the crucial step is the determination of a cutoff angular momentum $L_c$. 
  This is done 
  by calculating the complete fusion cross section within the sum-rule model,
  i.e., by using Eq.(\ref{SUMR}). We proceed as follows: the partial total fusion cross sections $\sigma_{\rm TF}^{(L)}$
  are summed from $L=0$ up to some angular momentum, such that the obtained sum
  fairly agrees with the experimental complete fusion data from Ref.\cite{Aguilera10}. Then such angular momentum
  that corresponds to this sum is taken as the cutoff angular momentum $L_c$ for a specific incident energy 
  $E_{\rm c.m}$.
  Within this procedure, we should allow a possible deviation in the determination of 
  $L_c$, with $\Delta L_c \sim \pm 0.5$.
  In Fig.~\ref{fig:3}, we notice that the calculated complete fusion cross section
  is in excellent agreement with the experimental data. The corresponding numerical values of $L_c$
  are summarized in Table~\ref{tab1}. Recalling that the angular momentum of  the relative center-of-mass motion was
  truncated at $500$, these numerical 
values serve to further confirm that indeed the complete fusion is a process that occurs at lower angular momenta, as
  long asserted in Ref.~\cite{Wik20}. 
  For example, by considering $E_{\rm c.m.}=37.5\,{\rm MeV}$, one can deduce that $L_c$ represents only about 3\% 
  of $L_{\rm max}$.  Another relevant aspect to be noticed in this table is the observed dependence
 of $L_c$ on the incident energy, 
 whereas the $L_{\rm crit}$ of Ref.~\cite{Wik20} is fixed for a specific reaction, as given for energies well above the 
 Coulomb barrier.  

\begin{table}
		\caption{For each incident energy, given in the first line, the second line provides the approximate values for the
		corresponding cutoff angular momenta $L_{c}$ (in units of $\hbar$), consistent with experimental data for 
		complete fusion cross sections, as given in Fig.~\ref{fig:1}. } 
		\label{tab1} \begin{center}
		\begin{tabular}{c|ccccccc}
			\hline\hline
			$E_{\rm c.m.}$(MeV)& 32.10 &33.00 & 33.90 & 34.80 & 35.70 & 36.60 & 37.50 \\
                          \hline
			$L_c $  & 9 & 11 & 12 & 13 & 14 & 15& 16 \\
			\hline\hline
		\end{tabular} \end{center}
\end{table} 
	
Having obtained the cutoff angular momentum $L_c$, considering available experimental data, let us now study the dependence
of the total fusion cross section on the variation of the projectile ground-state binding energy below the experimental
value for $L\le L_c$.
\begin{figure*}[!!t]
	\begin{center}
		\resizebox{150mm}{!}{\includegraphics{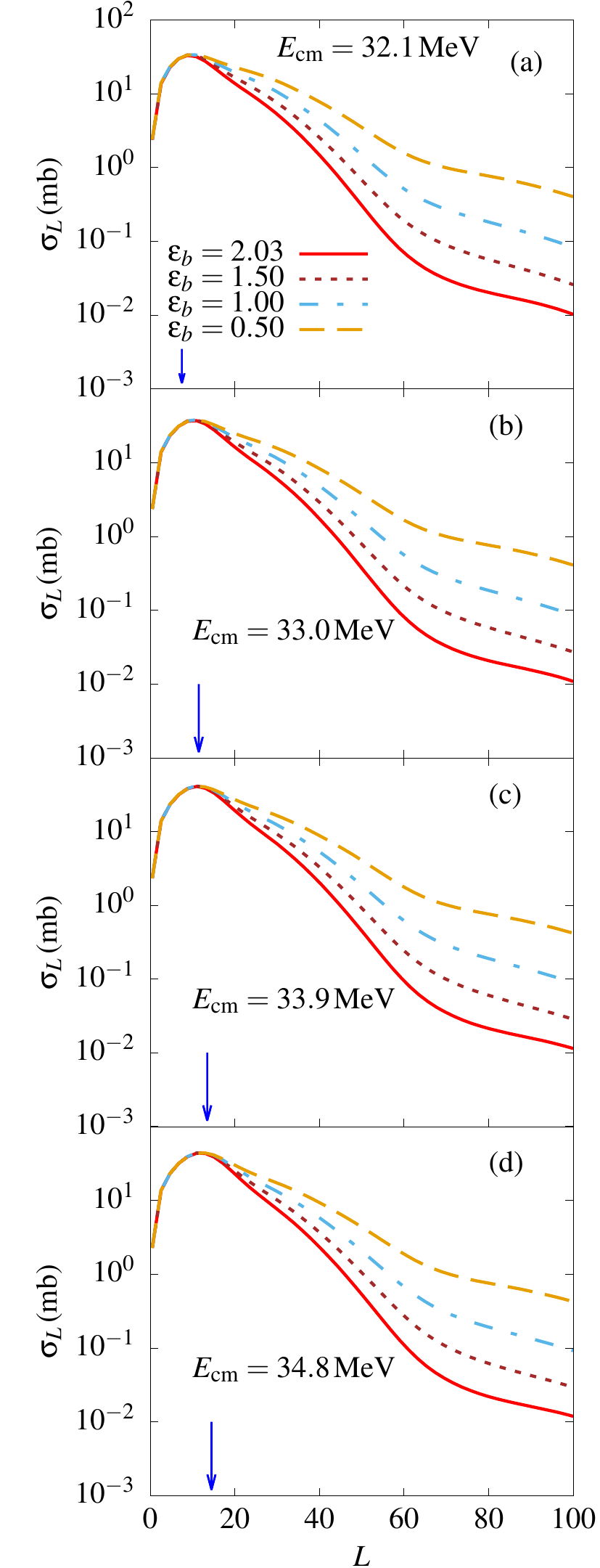}\includegraphics{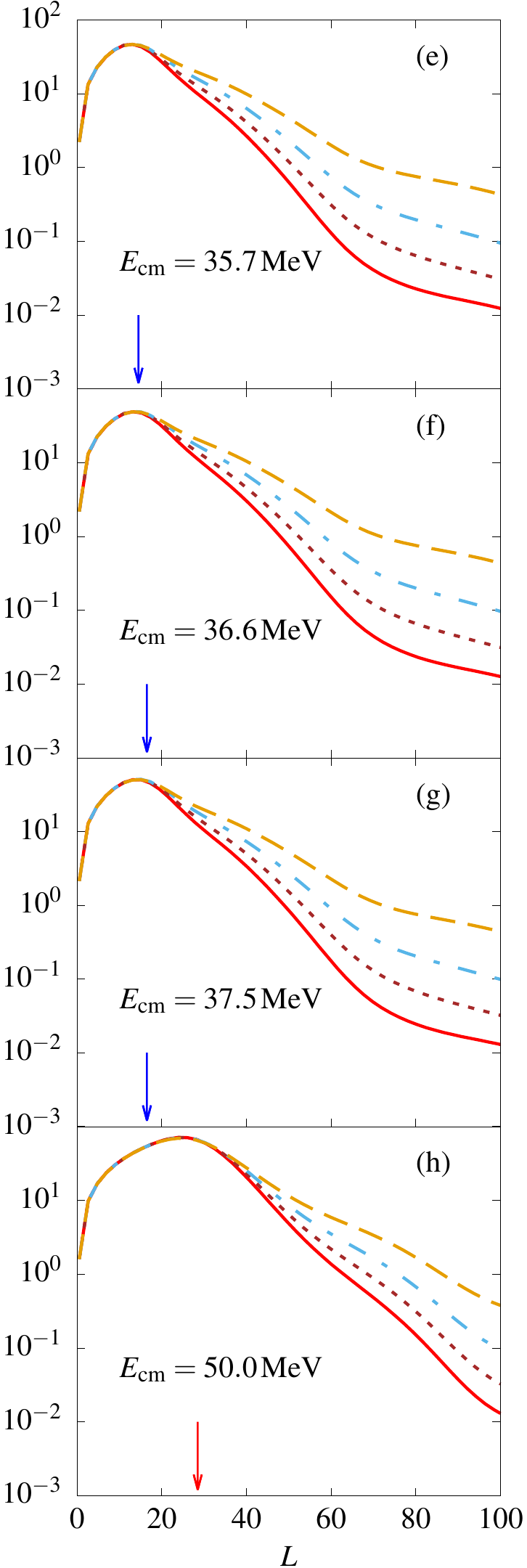}\includegraphics{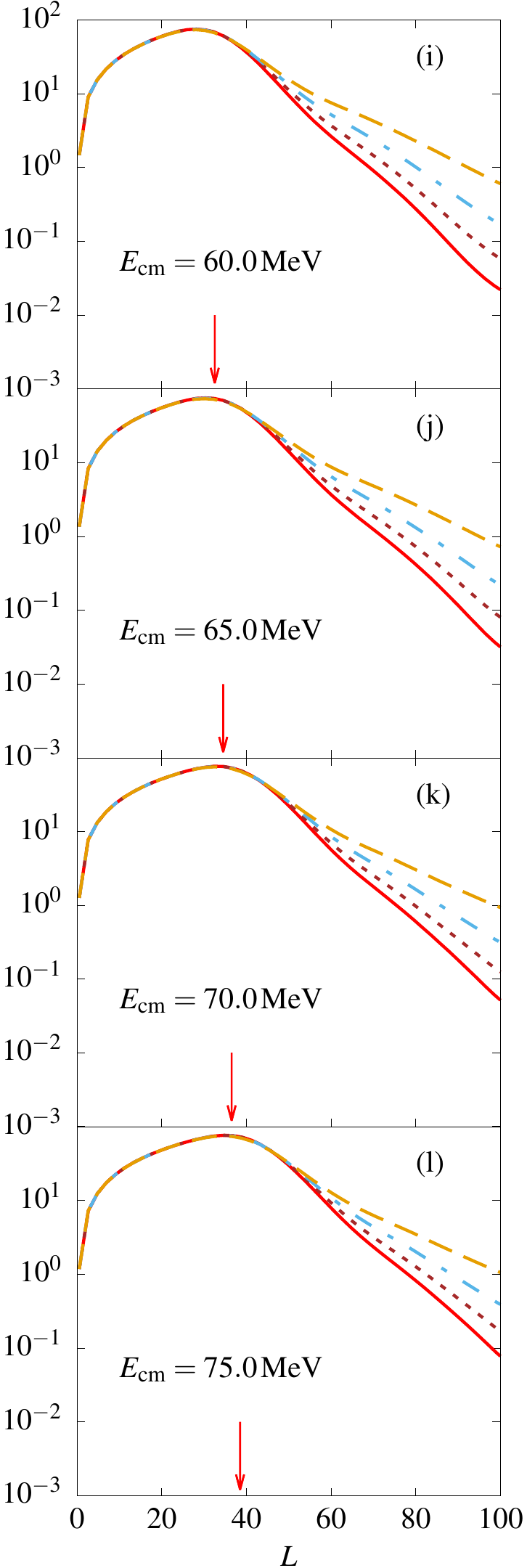}}
	\end{center}
	\caption{\label{fig:4}
	Angular momentum distributions for the total fusion cross sections (in units of $\hbar$), considering different projectile 
	binding energies $\varepsilon_b$ for $^8{\rm Li}$ projectile, as indicated in the frame (a). The results shown in the panels
	(a)-(g) are for the incident energies given in Table~\ref{tab1}, which are in agreement with experimental data, in case
	$\varepsilon_b=2.03$MeV.
	From our results obtained from calculation with incident energies not experimentally available, which are presented in 
	Table~\ref{tab2}, we include the panels (h)-(k).  The approximate values for the $L_c$ positions, which are presented 
	in the tables, are indicated by the arrows.
	}
\end{figure*}
To this end, the $L$-distribution total fusion cross sections, given by $\sigma_L\equiv \sigma_{\rm TF}^{(L)}$
are depicted in Fig.\ref{fig:4}.  
 The blue arrows in the panels (a)-(g) of this figure represent the values of $L_c$ corresponding to each incident energy as listed 
 in Table~\ref{tab1}. Looking carefully at this
figure, it is interesting to observe that the total fusion cross section has a rather insignificant dependence on the binding energy
for $ L\le L_c$, whereas it strongly depends on this energy for $L>L_c$, in
agreement with our assessment in the introduction. 
Consequently, the complete fusion cross section, which is defined in the $L\le L_c$ window, will exhibit a weaker dependence 
on the projectile ground-state binding energy. On the other hand, the
 incomplete fusion, which is defined in the $L>L_c$ window, will have a stronger dependence on this energy. 
 These results  show that $L_c$ can also be regarded as the maximum angular momentum such that, for $L\le L_c$, the total 
 fusion cross sections exhibit an insignificant dependence on variations of the projectile ground-state binding energy. 
 In fact, we have used this assessment to empirically determine $L_c$ for incident energies where 
 experimental data are not available. The computed results, considering several values for $E_{c.m.}$ from 45 to 75 MeV, are
 obtained by analyzing the corresponding behaviors of $\sigma_L$, for different projectile binding energies, which will indicate  
 $L_c$ as the approximate upper limit for $L$, such that the results become almost independent on the binding energies.  
 The panels (h)-(l) of Fig.~\ref{fig:4} are exemplifying how these approximate results given in Table~\ref{tab2} 
 are obtained, where the arrows (red) indicate the $L_c$ positions. 
 \begin{figure}[t]
	\begin{center}
	  \resizebox{8cm}{!}{\includegraphics{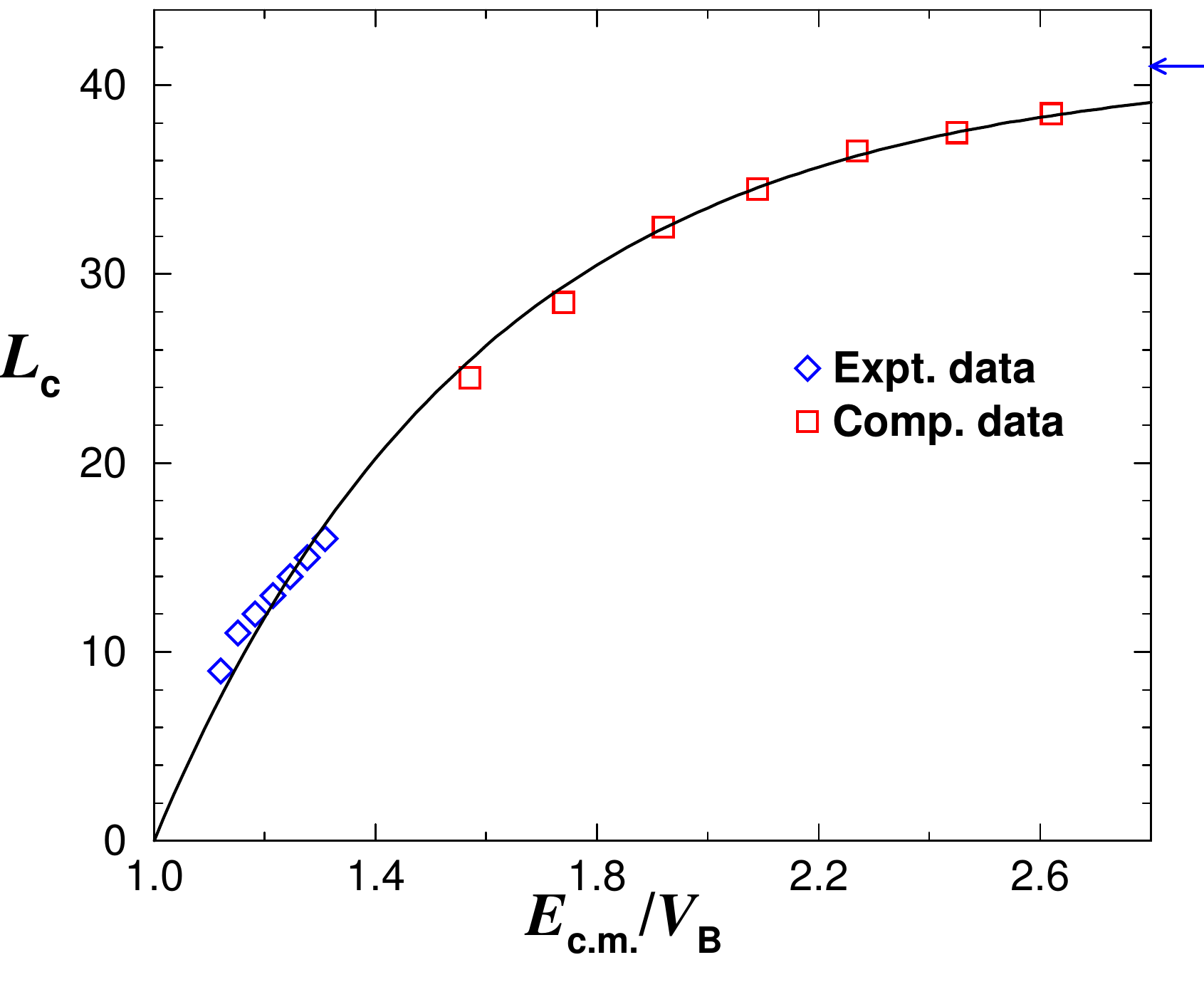}}
	\end{center}\vspace{-0.5cm}
	\caption{\label{fig:5} Cutoff angular momenta $L_{c} $ (in units of $\hbar$), are shown as a function of ${E_{\rm c.m.}/V_B}$, 
	where the solid curve is given by Eq.~(\ref{bestfit}). 
	The data points are those given in Tables~\ref{tab1} (diamond symbols) and \ref{tab2} (square symbols), from which one 
	should allow a possible error in the extracted $L_c$ of about $\pm 0.5$. The arrow is indicating the 
	corresponding sum-rule critical position for the $^8{\rm Li}+^{208}{\rm Pb}$ reaction, $L_{crit}\approx 41$.}
\end{figure}
   In order to search for a possible relationship between $L_c$ and $L_{\rm crit}$, we look for an expression to fit  
 the values presented in the Tables~\ref{tab1} and \ref{tab2}.  As shown in Fig.\ref{fig:5}, for colliding energies above the 
 Coulomb barrier $V_{\rm B}$, they can be well represented by the empirical expression 
\begin{eqnarray}\label{bestfit}
L_{c}\simeq \left\{1-\exp\left[-1.7\left({E_{\rm c.m.}/V_B}- 1\right)\right]\right\}  L_{\rm crit},
\end{eqnarray}
where the factor 1.7 in the exponential is an adjustable parameter.
In our approach, we assume $V_B=$28.65 MeV, from the S\~ao Paulo
Potential~\cite{Chamon20}, with $L_{\rm crit}$ being the well-known critical limit for the complete fusion
 according to the sum-rule model, which is obtained from the equilibrium condition of the Coulomb, nuclear and 
 centrifugal forces~\cite{Wik20,Wik10,1982Wik}:
\begin{equation}\label{lcrit}
\left(L_{\rm crit}+\frac{1}{2}\right)^2= \frac{\mu(R_p+R_t)^3}{\hbar^2}
\left[2\pi(\gamma_p+\gamma_t)\frac{R_pR_t}{R_p+R_t} 
-\frac{Z_pZ_t e^2}{(R_p+R_t)^2}\right],
\end{equation}
where $\gamma_i=0.95\left[1-1.78 ({1-2Z_i/A_i})^2\right] {\rm MeV} {\rm fm}^{-2}$ 
$(i\equiv p,t)$ are the surface tension coefficients, $R_{i}$ are the half-density nuclear radii, $\mu$ is the reduced mass, and 
$A_i$ the mass numbers. 
For the present $^8{\rm Li}+^{208}{\rm Pb}$ reaction, by assuming ($R_p, R_t)$ from Ref.~\cite{2013-Angeli}, with
$R_p=R_{^8{\rm Li}}=$2.339 fm and $R_t=R_{^{208}{\rm Pb}}=$ 5.501 fm, we obtain $L_{\rm crit}\approx 41$. 
The expression (\ref{bestfit}) is indicating that a more general relation may be found for the energy-dependent $L_c$ in terms 
of $L_{\rm crit}$, which requires further investigation to be well established.
 \begin{table}
   \begin{center}
     \caption{
For each incident energy, given in the first line, in the second line we present the corresponding approximate 
cutoff angular momenta $L_{c}$ (in units of $\hbar$), within uncertainty of about $1 \hbar$ given inside the 
parenthesis.
For $L\le L_c$, the total fusion cross section should exhibit an insignificant dependence on variations of 
$\varepsilon_b$.  The panels (h) to (l) of Fig.~\ref{fig:4} are exemplifying how the $L_c$ positions are obtained.
}  
		\label{tab2}
		\begin{tabular}{c|cccccccc}
			\hline\hline
			$E_{\rm c.m.}$(MeV)& 45 & 50      & 55   & 60   & 65    & 70  & 75    \\  \hline
			$L_c $       & (24-25) & (28-29) & (32-33) 
			& (34-35) & (36-37) & (37-38) & (38-39)\\
			\hline\hline
		\end{tabular} 
   \end{center}
 \end{table} 

\begin{figure}[!!h]
	\begin{center}
		 \resizebox{16cm}{!}{\includegraphics{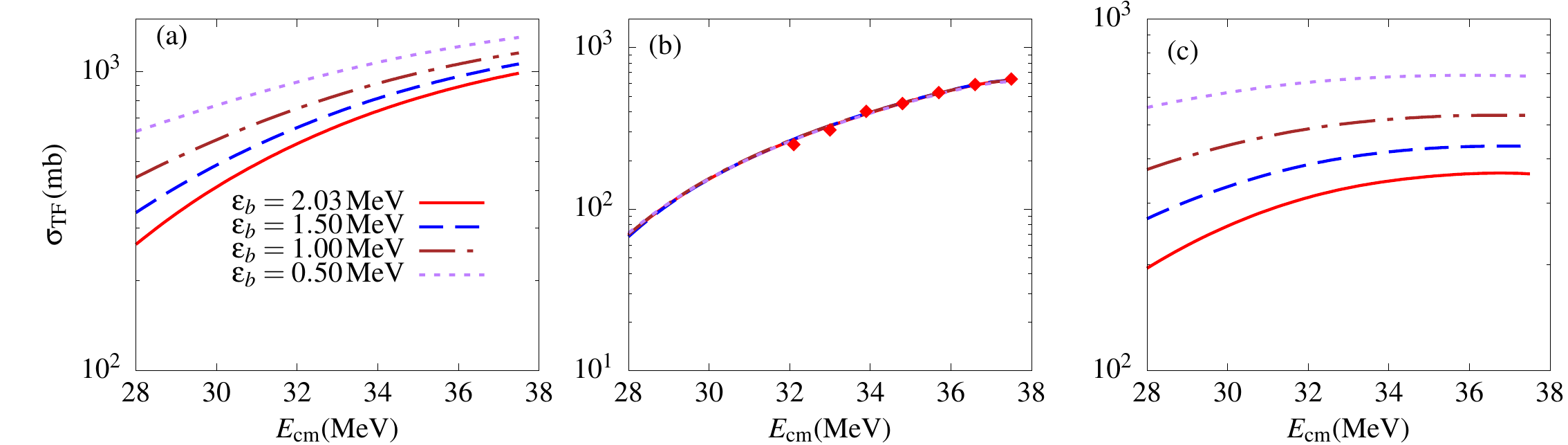}}		
	\end{center}
	\caption{\label{fig:6}
		Total, complete and incomplete fusion cross sections, given as functions of the $E_{c.m.}$,
		for different ground-state energies $\varepsilon_b$, as indicated inside the frame (a).
		The triangles in panel (b) refer to data results given in Table~\ref{tab1},
		obtained for $\varepsilon_b=2.03$MeV.}
\end{figure}

\begin{figure}[t]
\begin{center}
  \resizebox{14cm}{!}{\includegraphics{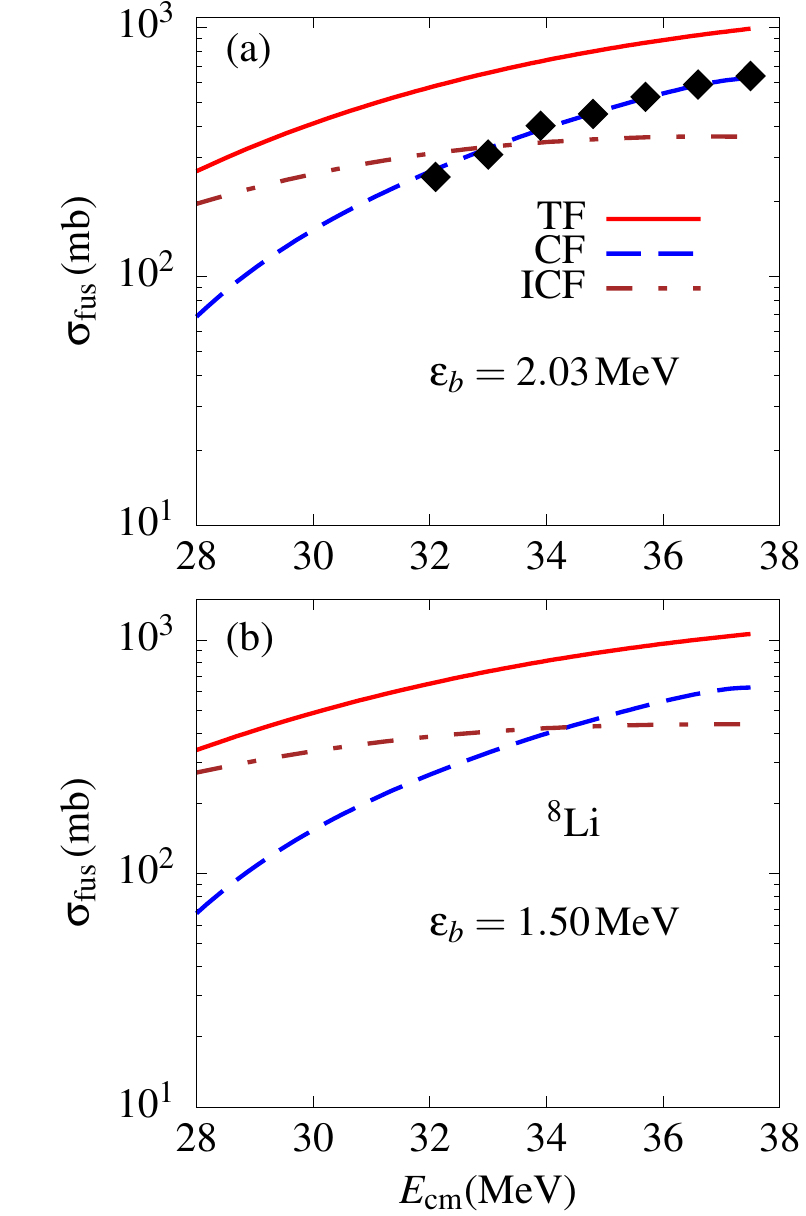}\includegraphics{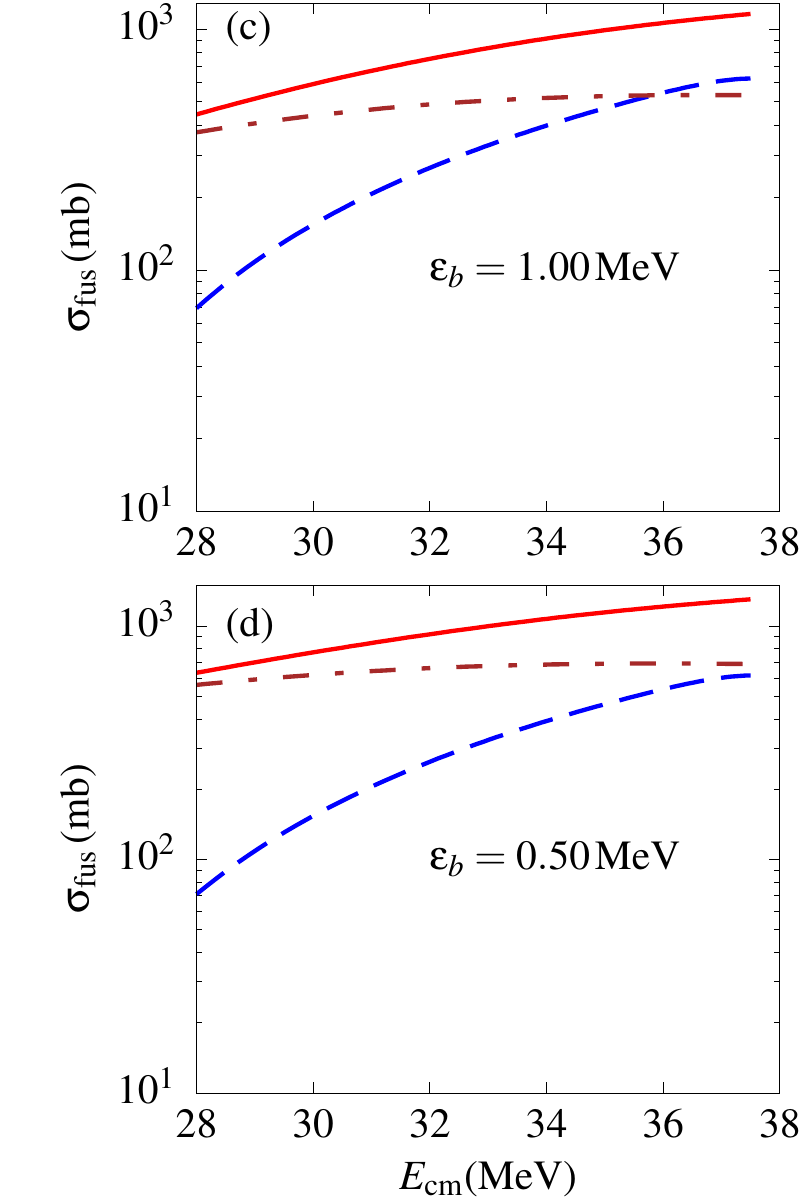}}
\end{center}\vspace{-0.5cm}
\caption{\label{fig:7}
Total (solid lines), complete (dashed lines) and incomplete (dot-dahed lines) fusion cross sections are represented
as  functions of  $E_{c.m.}$, for different ground-state binding energies $\varepsilon_b$ (as indicated inside the frames).
The panels are displaying the competition between complete and incomplete fusion cross sections.
The symbols presented in the panel (a) correspond to results given  in Table~\ref{tab1}. 
}
\end{figure}
Coming back to the dependence of total, complete and incomplete fusion cross sections on the projectile ground-state binding
energy, these cross sections are plotted in Fig.\ref{fig:6}, for the different four ground-state binding energies.
Indeed, the inspection of this figure shows that, the complete fusion cross section [Fig.~\ref{fig:6}(b)] has an 
insignificant dependence on the variation of the ground-state binding energy, in clear agreement 
with the results obtained for $^7{\rm Li}$ in Ref.\cite{Lei100}, while the incomplete fusion cross section [Fig.~\ref{fig:6}(c)]
is strongly dependent
on this energy. It follows that the $^7{\rm Li}$ and $^8{\rm Li}$ complete fusion cross sections are both
insignificantly dependent on the variation of  
ground-state energy below the respective experimental values,  although the former is modeled as a
cluster of alpha and triton nuclei. These results further indicate that breakups of 
these isotopes prior to reaching the fusion barrier alone is not a sufficient condition to explain the suppression of complete fusion, a
conclusion reported in Refs.~\cite{Cook20,Cook21}. 
{For completeness, Fig.\ref{fig:7} displays the competition between
the complete and incomplete fusion cross sections for the different binding energies considered. As expected, 
it is seen that, the incomplete fusion cross section is dominant over its complete counterpart at low incident energies, whereas
the complete fusion cross section prevails for larger incident energies [see Fig.~\ref{fig:7}(a)-(c)]. However, looking carefully at this
figure, it appears that as the binding energy decreases, the incomplete fusion cross section becomes dominant for all
incident energies within the interval studied here (around the Coulomb barrier)
[see Fig.~\ref{fig:7}(d)]. Another interesting aspect in this figure that better depicts the results in
Fig.\ref{fig:6} (b) is that all the different considered binding energies provide an excellent fit of the experimental data.}
\begin{figure}[t]
\begin{center}
  \resizebox{14cm}{!}{\includegraphics{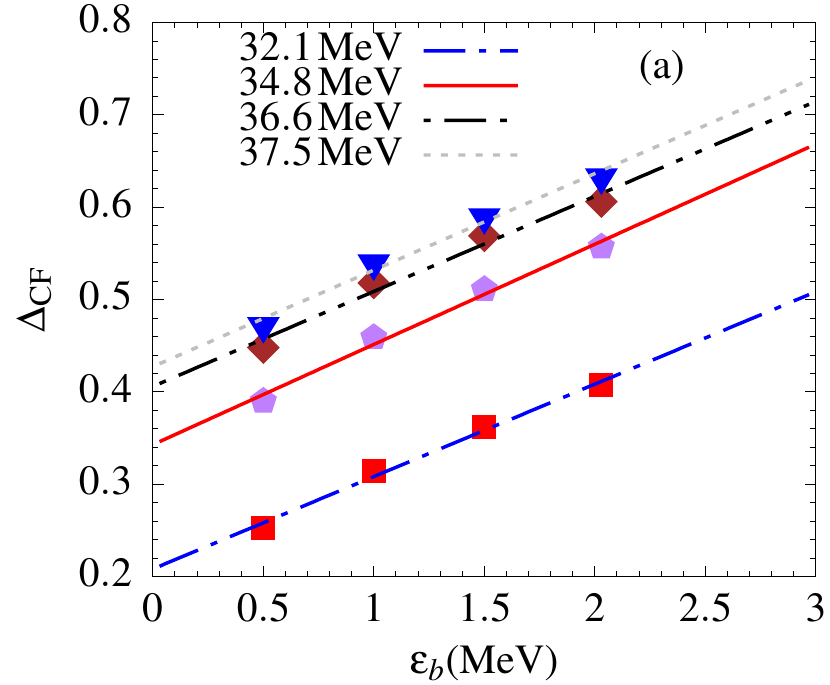}\includegraphics{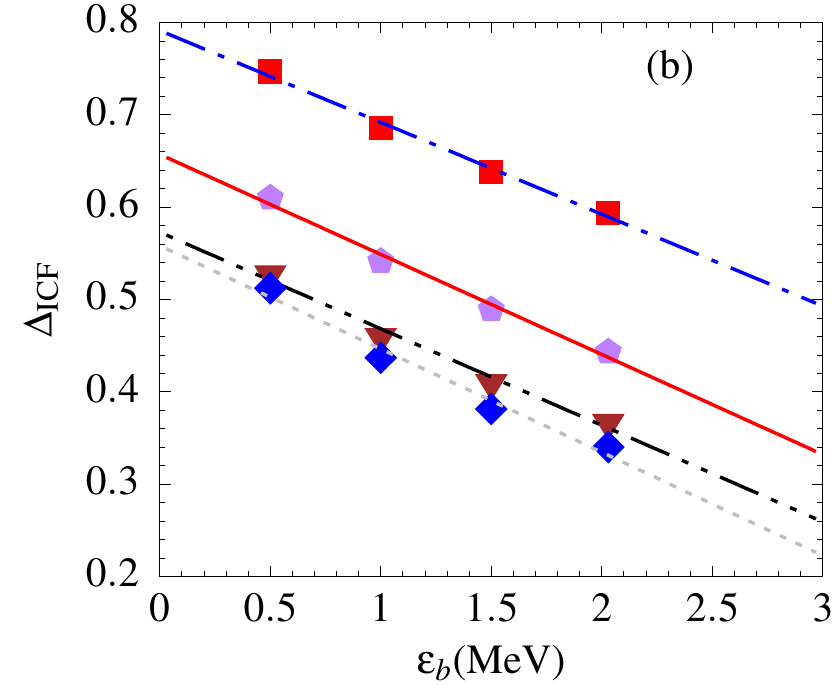}}
\end{center}\vspace{-0.5cm}
\caption{\label{fig:8} Complete fusion suppression [panel (a)], with the corresponding incomplete fusion enhancement [panel (b)]
  as functions of the projectile ground-state binding energy, $\varepsilon_b$ (in MeV), for a given set of incident energies
  indicated inside the panel (a).}
  \end{figure}

Since the complete fusion is verified to be independent on
variations of the binding energy below the experimental value, it is
important to investigate how 
is it suppressed owing to the variation of the binding energy.
It has been shown that the complete fusion in the $^8{\rm Li}+{}^{208}{\rm Pb}$, $^{209}{\rm Bi}$ reactions, is
suppressed by about 30\% \cite{Cook20,Aguilera10}, although the $^8$Li projectile breaks up into one charged and one uncharged 
fragments. As the complete fusion suppression can be analyzed through its contribution to the total fusion cross section, we first define 
 the complete ($\Delta_{\rm CF}$) and incomplete ($\Delta_{\rm ICF}$) contributions to the total $\sigma_{\rm TF}$ by 
\begin{eqnarray}\label{contributions}
  \Delta_{\rm CF}=1-\left(\frac{\sigma_{\rm ICF}}{\sigma_{\rm TF}}\right),\quad
  \Delta_{\rm ICF}=\left(1-\frac{\sigma_{\rm CF}}{\sigma_{\rm TF}}\right).
\end{eqnarray}
The corresponding results for complete and incomplete fusions are plotted in the panels (a) and (b) of  Fig.~\ref{fig:8}, 
respectively.  In these two panels, the results are plotted as functions of binding energies 
$\varepsilon_b$.  By inspecting this figure, one observes that the complete fusion is well-suppressed owing to the 
decrease of the binding energy [Fig.\ref{fig:8}(a)], while the incomplete is enhanced [Fig.\ref{fig:8}(b)].
In particular, a careful look it this figure shows that for $E_{\rm c.m.}=37.5\,{\rm MeV}$ the complete fusion accounts for 
over 60\% of the total fusion cross section for $\varepsilon_b=2.03\,{\rm MeV}$, which represents a suppression of about 35\% 
which fairly agrees with suppression factor reported in  
Refs.~\cite{Cook20,Aguilera10}. 
As observed, this contribution drops significantly as $\varepsilon_b$ decreases, and it is about 25\% (which represents a 
suppression of about 75\%) for $\varepsilon_b=$0.50 MeV. It is clear from Eq.(\ref{contributions}),
that the suppression of the complete fusion owing to the decrease of the projectile ground-state binding energy,
is due to a stronger dependence of the incomplete fusion on this energy.

It has been shown that the incomplete fusion contribution, which is often referred to as incomplete fusion probability, 
decreases linearly as the projectile ground-state binding energy increases, see for example Ref.~\cite{Hinde2010}. 
Generally, this analysis is done by considering different projectiles with different binding energies, atomic masses, as 
well as ground-state configurations.
Since that $^8$Li is primarily formed by $^4$He$+^3$H$+n$ (as studied in Ref.~\cite{1998Grigorenko}),  
it should also be stressed that after the breakup the ICF process may involve fusion with charged clusters $^4$He$+^3$H. 
Therefore, in order to better elucidate the ICF for the reaction under study, a four-body CDCC calculation 
would be helpful, with the $^8$Li  being treated within a three-body configuration.
Or , within a sequential breakup, in which the  $^8$Li first breaks into $n+^7$Li, followed by a  second step 
with the $^7$Li split into $^4$He$+^3$H. 
Hence, the results obtained may also depend on different factors other than the binding energy. 
However, it is interesting to observe in Fig.\ref{fig:8}(b) a perfect linear dependence of the ICF contribution on the 
ground-state binding energy for all the incident energies; a linearity which is also displayed in Fig.\ref{fig:8}(a) 
for the corresponding complementary complete fusion contribution.

\section{Conclusions}
\label{conclusion}
We have investigated in some detail the dependence of the total, complete and incomplete fusion cross sections 
on variations of the projectile ground-state binding energy below the experimental value for the
$^{8}{\rm Li}+{}^{208}{\rm Pb}$ reaction.  By adopting the sum-rule model of Refs.\cite{Wik10,Wik20}, 
and by using the complete fusion experimental data, an angular momentum cutoff $L_c$ is determined, 
such that the complete fusion cross section occurs at angular momenta ($L$) lower than $L_c$ ($L\le L_c$). 
By fitting the $L_c$ numerical values, we have obtained an expression linking both $L_c$ and the well-known 
critical value $L_{\rm crit}$, which was derived in Ref.~\cite{Wik20}.
The expression for $L_{\rm c}$ is  a function of the  incident energy $E_{\rm c.m.}$ above the Coulomb barrier, 
whereas $L_{\rm crit}$ is an energy independent parameter.
  Therefore, the results obtained hint to a possibility of extending the sum-rule model to energies around the 
  Coulomb barrier.  On the other hand, our findings indicate that a combination of the CDCC and sum-rule 
  models can provide a better description of complete fusion processes, which is pointing out to an interesting   
extension of this study by considering incident energies below the Coulomb barrier.

The study of the dependence of the total fusion cross section on variations of the binding energy $\varepsilon_b$  
shows that the total fusion cross section has an insignificant dependence on such variations for angular momenta 
in the limit for $L\le L_c$, while it strongly depends on these variations for $L>L_c$. 
Consequently, the complete fusion, which is defined in the $L\le L_c$ angular momentum window, is found to exhibit an
insignificant dependence on $\varepsilon_b$ variations, whereas the incomplete fusion, defined in the $L>L_c$ 
window, is strongly dependent on $\varepsilon_b$.  This is one of the main outcomes of this study.
It follows that weak binding energy (or breakup prior to reaching the fusion barrier) alone, is not a sufficient 
ingredient to explain the suppression of complete fusion, as also it was concluded in Refs.\cite{Cook20,Cook21}. 
Therefore, the results in the present work can be regarded as a further theoretical support to the conclusions 
drawn in these references. Nevertheless, the complete fusion cross section is found to be stronger suppressed 
owing to the decrease of the binding energy. This comes rather from a strong enhancement of the incomplete
fusion cross section due to the decreasing of the binding energy. 
We have also verified that the complementary complete and incomplete fusion contributions present approximate 
linear dependences on variations of the ground-state binding energy, when considering the available 
incident-energy experimental results.

\section*{Acknowledgements}
B.M. is grateful to the South American Institute of Fundamental Research (ICTP-SAIFR) for local facilities. 
We also thank the following agencies for partial support: 
Conselho Nacional de Desenvolvimento Cient\'\i fico e Tecnol\'ogico [INCT-FNA Proc.464898/2014-5 (LT and JL), 
Proc. 306191-2014-8(LT)], and Funda\c c\~ao de Amparo \`a Pesquisa do Estado de S\~ao Paulo [Projs. 2017/05660-0(LT)].
\\

\end{document}